\documentclass{ifacconf}

\usepackage{graphicx}      
\usepackage{natbib}        
\usepackage{balance}
\usepackage{footnote}
\usepackage{url}

\begin{document}
\begin{frontmatter}

\title{Strategies for Big Data Analytics through Lambda Architectures in Volatile Environments}


\author[First]{Veith, Alexandre da Silva}
\author[First]{Anjos, Julio C. S. dos}  
\author[First]{de Freitas, Edison Pignaton}
\author[Second]{Lampoltshammer, Thomas J.}
\author[First]{Geyer, Claudio F.}

\address[First]{Federal University of Rio Grande do Sul (UFRGS), 
   Porto Alegre -RS - P.O.Box 15064 - Brazil (e-mail: alexandre.veith@ufrgs.br and \{jcsanjos, edison.pignaton, geyer\}@inf.ufrgs.br)}
\address[Second]{Danube University Krems, Department for E-Governance and Administration, Dr.-Karl-Dorrek-Str. 30, 3500 Krems, Austria\\ (e-mail: thomas.lampoltshammer@donau-uni.ac.at)}

\begin{abstract}                

Expectations regarding the future growth of Internet of Things (IoT)-related technologies are high. These expectations require the realization of a sustainable general purpose application framework that is capable to handle these kind of environments with their complexity in terms of heterogeneity and volatility. The paradigm of the Lambda architecture features key characteristics (such as, robustness, fault tolerance, scalability, generalization, extensibility, ad-hoc queries, minimal maintenance, and low-latency reads and updates) to cope with this complexity. The paper at hand suggest a basic set of strategies to handle the arising challenges regarding the volatility, heterogeneity, and desired low latency execution by reducing the overall system timing (scheduling, execution, monitoring, and faults recovery) as well as possible faults (churn, no answers to executions). The proposed strategies make use of services such as migration, replication, MapReduce simulation, and combined processing methods (batch- and streaming-based). Via these services, a distribution of tasks for the best balance of computational resources is achieved, while monitoring and management can be performed asynchronously in the background.  

\end{abstract}

\begin{keyword}
Internet of Things (IoT), Scheduling, Batch processing, Stream processing, Cloud computing, Grid computing
\end{keyword}

\end{frontmatter}
\section{Introduction}

The  concept of the Internet  of  Things (IoT) can be described as the seamless fusion of virtual environments and contained objects with their real-world counterparts \citep{2011:Uckelmann}. In return, this makes the creation of robust, flexible, and dynamic applications imperative, in order to handle heterogeneous and volatile environments.  A major  aspect in this regard is represented by the challenge to handle vast amounts of data, including all relevant processing steps, in particular data analytics. Due to this fact, the area of big data analytics  has  attracted  high levels of attention  of   industry  and  academia alike. This fact is represented by the total increase of data-driven projects by 125\% during the period 2014-2015.\footnote{IDG - \url{http://www.idgenterprise.com/}} So far, the majority of big data deployments were initially using batch processing-oriented approaches (i.e. the entire amount of data is gathered, stored, and afterwards processed step-by-step) \citep{2014:hu}. However, \emph{batch processing} has no support for low-latency scenarios. Thus, a new model called \emph{stream processing} or \emph{oriented-to-events processing} has witnessed a huge increase in volume and availability \citep{2014:tudoran}. The handled events are usually characterized by a small unit size (in the order of kilobytes), but have overwhelming collection rates, due to the continuous data flow. To overcome this issue, new \emph{stream processing} frameworks have emerged, such as Apache Storm, Spark, Flink or S4.






Over time, \emph{stream processing} and its associated processing engines have evolved up to the point of the introduction of the Lambda Architecture (LA) paradigm \citep{2013:Marz}. Lambda Architectures are designed to handle vast amounts of data in conjunction with both \emph{batch} and \emph{stream processing} methods. While \emph{batch processing} helps to reduce latency, to improve data transfer, to provide fault-tolerance, as well as a comprehensive and accurate view upon the data, \emph{stream processing} provides capabilities to deal with real-time data. Thus, the rise of LAs  are directly related to the rapid growth of Big Data real-time analytics.

According to \cite{2013:Ewen}, the Lambda Architecture paradigm is the starting point of the so-called \emph{$4^{th}$ Generation of Data Processing Engines} that comprise several features regarding the design and implementation of processing engines for massive data, such as robustness, fault tolerance, low latency of reading and updating, scalability, generalization, extensibility, ad-hoc queries, and minimal maintenance. To achieve these properties, the architecture foresees a Big Data system to be constructed in several layers. \cite{2015:Anjos} presented the SMART platform, which is a modular framework for Big Data analysis. SMART considers a large variety of data sources, such as distributed datasets and social networks, where there is a clear need for standardization. The Dispatcher module (DM) in the SMART platform is an orchestration system that needs several policies to the managing data and tasks. This paper aims at the improvement of the decision-making engine of the Dispatcher module based on the computational capacity of the machines via scheduling strategies and advanced execution setups regarding data streams to achieve an optimal distribution of tasks for the best balance of computational resources.



This paper is structured as follows: Section \ref{sec:state_of_the_art} sets out the state-of-the-art for data-intensive computation, establishes the framework in this landscape, compares it to others and demonstrates how it works in relation to others considering heterogeneous infrastructures, hybrid infrastructures, and hybrid engines. 
Section \ref{sec:architecture_overview} presents the SMART platform and its main modules. Section \ref{sec:dispatcher_module} details the Dispatcher Module and discusses the strategies required to overcome the environmental limitations of this module. Section \ref{sec:conclusion} concludes the paper and reports opportunities for future work. 

\section{Related Work}
\label{sec:state_of_the_art}

\subsection{Heterogeneous Infrastructures}
\label{subsec:heterogeneous_infrastructures}
\emph{JetStream} is a set of strategies for efficient transfers of events between cloud data centers \citep{2014:tudoran}. \emph{JetStream} is self-adapting regarding streaming conditions. It aggregates the available bandwidth and enables the routing of data through cloud sites. The study by \cite{2014:tudoran} focuses on event transfers between inter- and intra-nodes. The authors propose an adaptive cloud batching in form of an algorithm that aggregates the streams in batches, resulting in latency reduction. However, it just considers the latency and not the volatility. The work concentrates on environments where the computational resources can break away unexpectedly and the scheduling policies must be adapted to it. 


SMART \citep{2015:Anjos} is a platform that offers an efficient architecture for Big Data analysis applications for small and medium-sized organizations. Its implementation considers heterogeneous data sources and aims at data analysis scenarios in geo-distributed environments, considers cost, fault tolerance, network overhead, I/O throughput, as well as the minimization of data transfers between computational resources. Yet, these parameters are not enough to work with volatile environments, especially regarding \emph{stream processing}. To overcome these environment limitations, it is necessary incorporate information from physical components, such as memory, CPU speed, and storage. Thus, the overall capacity impacts the overall performance. In addition, regarding the volatility, replication must be incorporated as a fault control mechanism.

Similarly, \cite{Pham:2016} purpose a generic, extensible, scalable, fine-grained, and re-configurable multi-cloud framework. It is based on a lightweight kernel and provides a hierarchical Domain Specific Language (DSL). The DSL allows for a fine-grained level of administration. However, the proposed solution does not control the workload at the nodes and possible faults, as the Deployment Manager is just an interface to set the devices and the Virtual Machines (VMs).

\subsection{Hybrid Infrastructures}
\label{subsec:volatile_infrastructures}
\emph{BIGhybrid} summarizes the main features of a Hybrid MR environment based on the merge of two environments, namely a Cloud (MR-BlobSeer) environment and a Desktop Grid (BitDew-MapReduce) environment \citep{Anjos:2016}. The Global Dispatcher located outside the Desktop Grid (DG) as well as of the cloud environment features middleware functionality for handling task assignments and input data from users. It is a distributed data storage system that manages policies for data splitting and distribution in batch applications such as MapReduce. The working principle is similar to the publish/subscribe service, where the system obtains data and publishes the computed results. This approach has several drawbacks, if applied to \emph{stream processing}, due to delays regarding the processing of responses.

\emph{HybridMR} is a model for the execution of MapReduce on hybrid computation environments (Cloud and DG) developed by \cite{2015:Tang}. Two innovative solutions are proposed to enable such large-scale data-intensive computation: (i) HybridDFS, which is a hybrid distributed file system. HybridDFS features reliable distributed storage that alleviates the volatility of desktop PCs (i.e., fault tolerance and file replication mechanism); and (ii) a Node priority-based fair scheduling (NPBFS) algorithm has been developed to achieve both data storage balance and job assignment balance by assigning each node a priority through quantifying CPU speed, memory size, as well as input and output capacity. The NPBFS approach is very interesting because it uses some miscellaneous environment variables to schedule the tasks. Although regarding \emph{stream processing}, its application is not possible due to the high flow latency. \emph{HybridMR} just uses the flow rate to deploy the tasks, however, the rate does not consider particular task information.



\subsection{Hybrid Engines}
\label{subsec:hybrid_engines}
\emph{Apache Spark} is a framework introduced by \cite{2012:Zaharia} that uses resilient distributed datasets (RDDs) and enables efficient data reuse in a broad range of applications. RDDs are fault-tolerant, parallel data structures that are designed to allow users to keep intermediate results in memory, control their partitioning to optimize data placement, and manipulate them through a valuable set of operators. \cite{2015:Liao} presented some scheduling inefficiencies related to the time window that constructs the RDD. The batch interval needs to be dynamically adjusted, so that fluctuations within the data rate can be handled in a production environment and the total delay of every event can be controlled within a certain range for real-time scenarios. 

\emph{Apache Flink}, initially developed by \cite{Alexandrov:2014}, enables massively parallel \textit{in-situ} data analytics, using a programming model based on second order functions. Today, \emph{Apache Flink} is the state-of-the-art processing engine according to \cite{2013:Ewen}. The scheduling policies are designed to work at commodities environments (i.e., clusters and cloud). Commodities strategies do not work well with dynamic environments, as argued by \cite{2015:Peng} and \cite{2016:Eskandari}. It is essential for heterogeneous computing resources to acknowledge the surrounding environment.

\emph{Summingbird} by \cite{2014:Boykin} integrates batch and online analysis with the aid of a hybrid processing model, where access can be provided efficiently and seamlessly for aggregations across of long time spans while maintaining up-to-date values with a minimal latency. However, \emph{Summingbird} does not provide access to the Message Queue writing in Hadoop, it only has knowledge of that has been recorded. The scheduling policies are abstracted and the Hadoop and Storm systems handle the management.


This section has presented actual research opportunities in relation to the combination of volatile and heterogeneous environments, as well as scheduling improvements. Currently, \emph{stream processing} only has been performed in heterogeneous environments, while \emph{batch processing} was reserved for volatile environments. Overall, processing engines do not deal well with heterogeneity and volatility. In fact, most engines were designed to run in clusters and cloud computing environment. This fact makes a Round Robin solution attractive for the scheduling policy. Nonetheless, in complex environments such as Desktop Grid it is not possible to efficiently employ it. The restrictions are related to the task and node heterogeneity, as well as to network bandwidth. 

\section{SMART Architecture Overview}
\label{sec:architecture_overview}
\begin{figure}[ht]
	\centering
	\includegraphics[width=.45 \textwidth]{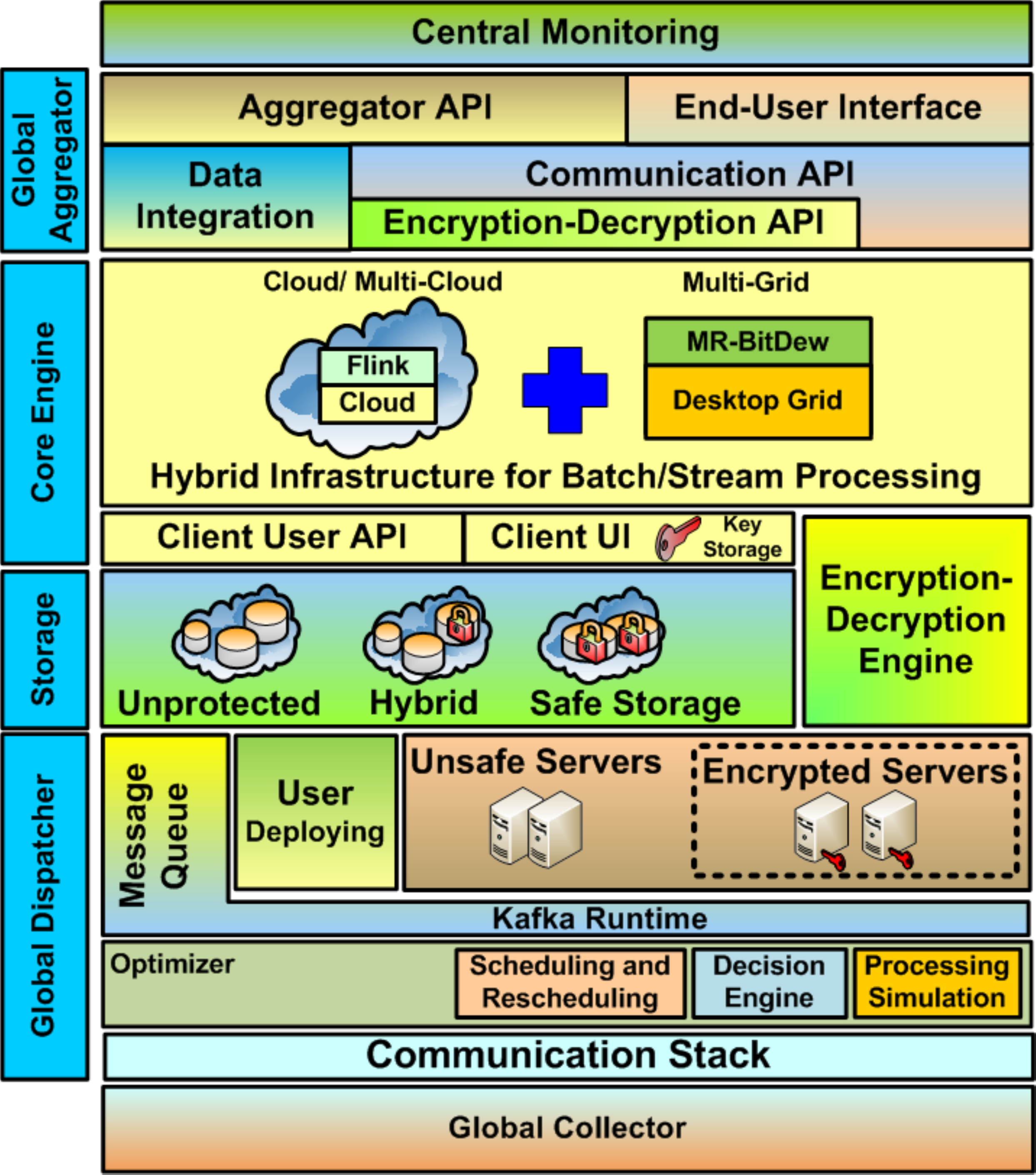}
	\caption{SMART Platform Overview}
        \label{fig:exampleFig1}
\end{figure}

The following section provides a brief overview of the layered architecture that is required to deploy a SMART-based environment. Figure \ref{fig:exampleFig1} presents the required four main modules: Global Collector, Global Dispatcher, Core Engine, and Global Aggregator.

The \emph{Global Collector} layer handles the management and coordination of the sensing modules. It is responsible for obtaining data from several sources and maintaining the data integrity mechanisms. The data integrity mechanisms filter possible noises in the hardware devices. The data is collected and serialized under a standard TCP/IP, which forms the communication stack for the \emph{Global Dispatcher}.

In the \emph{Global Dispatcher}, the data is decoupled from the lower layers in the message queue mechanism. It is put in a FIFO queue so that it can be distributed to severs in accordance to the availability of their resources in both \emph{Cloud/Multi-Cloud} and \emph{Grid/Multi-Grid} environments. The \emph{optimization} layer analyses the volume of input data and employs the \emph{Decision Engine} to make decisions about scheduling tasks and data through distinct environments. A simulation process implements an execution time prediction that will be used by the \emph{Decision Engine} to improve the accuracy of the scheduling mechanism. 

The \emph{Core Engine} must support hybrid systems, \emph{i.e.}, provision of streaming and batch computations at the same time. Thus, the Flink framework is an potential system that is worth considering. The MR-BitDew is another framework, which can improve computational performance, with the use of Volunteer Computing (VC) in a hybrid infrastructure. SMART allows computational resources to be taken from Cloud/Multi-Cloud, and Multi-Grid environments. A \emph{Client User API} provides an easy method for users to submit their applications and indicate the data sources. The \emph{Client UI} is a security interface that provides a single-user identification through an encrypted key. A key which is keep by the users is employed in the \emph{Encryption-Decryption Engine} to ensure the data is safe.
The intermediate results, processed in the \emph{Core Engine}, are serialized for the \emph{Global Aggregator} that must carry out the data consolidation. The \emph{Data Integration} module supports the data integrity and data integration. The last phase of the data processing is designed to generate an iterative execution and provide the results of the consolidation. A \emph{Communication API} is necessary to integrate the workers into a virtual network of data computation. 

The \emph{Global Aggregator} is a module that orchestrates the results of the aggregation and maintains the safety data mechanism for the end-users. The \emph{End-User Interface} shows the information in a user-friendly way through a \emph{Central Monitoring} system.

\section{Dispatcher Module Deployment Strategies}
\label{sec:dispatcher_module}
It is the main goal of this paper to overcome existing issues of the Global Dispatcher (GD) regarding data distribution in real time applications, via solutions presented by \cite{veith:2015}. This means that the Global Dispatcher will carry out the tasks of load balancing, and latency control (data stream processing and network bandwidth), provision of scalability, while reducing the costs for improvement of availability of resources. Additionally, it will monitor environmental behavior (network bandwidth, availability of processing capacity, time spent on completing tasks, churn and so on) for the Decision Engine. The role of the Decision Engine is to select the computing resources needed for carrying out a task. The task definition will be achieved by simulation, which will use BIGhybrid, and at the same time, to evaluate the environment for the re-scheduling processes (data placement and data movement). According to \cite{2012:Delamare}, it is possible to achieve QoS in the GD by monitoring the task execution and dynamically provisioning external, stable, and powerful Cloud resources to support the GD. They employed different strategies (i.e, Completion Threshold, Assignment Threshold, and Execution Variance) to control the Cloud usage.

Similarly, the model will employ strategies to control the use of the environment. To this end, it will have control over the execution of the system jobs. First of all, it will asynchronously create batch views asynchronously in the background of the tasks and storage of the data nodes (logs). Then, it will create the stream views, in which the latest logs (nodes and tasks) will be collected for the control. The combination of views will achieve a performance gain due to the fact that most of the information required will already have been generated when it is needed.

The clustering data method and processing in small batches will be used to obtain low latency for the stream-based processing (\cite{2014:das}). However, this method will be validated by conducting comparative studies to define the size of the batches and to determine if it will be the best technique for stream processing. This will involve validating the particular way of batching the streams (i.e., time-based, event-based, or hybrid form). Through the batch-sizing of \emph{stream processing}, it will reduce the latency, make it easier for the processing flow (i.e., by processing simulations) and facilitate the scheduling and rescheduling of tasks and data. This feature will overcome the BIGhybrid simulator problem (it cannot work with streams because of the SimGrid restrictions).

\subsection{Decision Strategies} 
\label{subsec:decision_strategies}
The Decision Engine (DE) will make the decision about scheduling and re-scheduling and thus will be the main component in the Global Dispatcher. It will combine environmental data (task simulation, availability of resources, network availability, costs, aggregation time and so on) with the strategies (i.e. batch and stream) to decide where to allocate the task to (node or Cloud).

\begin{figure}[htp]
\centering
\includegraphics[width=.48 \textwidth]{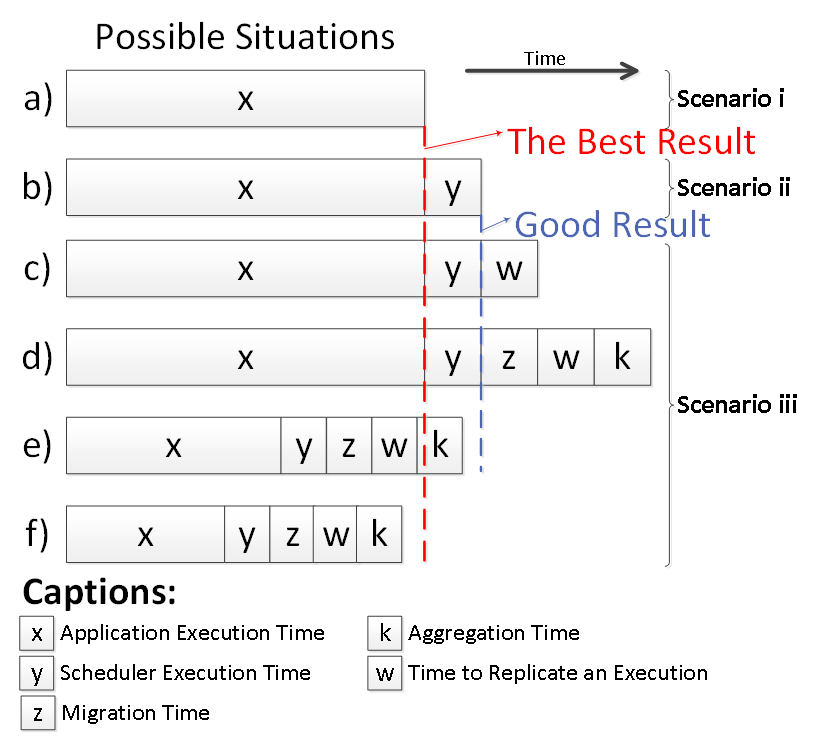}
\caption{Target Scenarios}
\label{fig:decision_engine_goal}
\end{figure}

Figure ~\ref{fig:decision_engine_goal} depicts the target scenarios, based on the type of strategies intended to obtain a Good or Best result in the load balancing. Differentials such as migration, elasticity, scheduling strategies, data aggregation and replication are highlighted in the model. Scenario i (part “a”, only the execution time) is the application (performing a set of tasks) without migration time, scheduling time, aggregation time or replication time. Scenario ii (part “b”) is the execution of the application and the scheduling (only to obtain the scheduling overhead). Scenario iii (parts “c”, “d”, “e” and “f”) involves the execution of all the services (migration, aggregation, scheduling and replication). In the part “c”, the migrations services are restricted, but still continue with the replication. In “d”, all the services will be operating, but with a strong impact on the time. “e”, the model will show the performance that is embodied in the strategies. If the time is lower than “b” this yields a Good Result. The “f” is the best scenario. The strategies at the scheduler reveal the best computing resources for execution. In this scenario, the performance will be greater than the “a” part, which is just the application of the execution (The Best Result).

Some factors must be included in the strategies to overcome the problem of time constraints in the system and achieve the Best Result or a Good Result. These include the ‘time to movement’ data and the task migrations, ‘time to data’ aggregate when the data is shared between a) the computing resources, b) time estimated to execute a task, c) data placement and e) computing resource rating. In this way the overall execution time can be reduced and allow a simple fail control (volatile).

The reasons for applying this feature to the architecture are set out below:
\begin{itemize}
\item \textbf{Migration}: This generates considerable gains in dynamic, heterogeneous and large- scale environments. Since there is a dynamic, the nodes might be in or out of the network (i.e., churn). This means that a certain task may be at an overload or a slow node, and there might be a machine that runs in less time (when the movements are counted). In this case, it is worth migrating the task;

\item \textbf{Aggregation}: Distributing the tasks belongs to the network and forms a part of a set; thus it will be necessary to group the results. The distribution will be able to obtain time and make use of the idle resources;

\item \textbf{Replication}: Replicating the task belongs to the network and means that it will be necessary to ensure the correct execution and reduce the time when a fault occurs. A fault generally occurs because of the volatile environment (Desktop Grid);

\item \textbf{Computing Resource Rating}: The rate will evaluate the resource data, network data and past execution data. This method assists in recognizing a good worker;

\item \textbf{Time Estimation}: Knowing the time before executing a task will make it easier to ensure a correct scheduling;

\end{itemize}

As discussed earlier in this paper, this work includes controls for monitoring and to provide decisions that will run asynchronously in the background. These controls combined to recent information (stream views) and historical information (batch views), will allow a hybrid control system. The result will enable settlements to be made safer and faster. In light of this, the following methods are combined:

\begin{itemize}
\item \textbf{Batch-Based}: The raw data stored in a distributed file system will be processed in a \emph{batch processing} way; 

\item \textbf{Stream-Based}: The recent data will be performed in near real-time fashion;
\end{itemize}

The combination of batch and stream strategies will be able to reduce the Dispatcher time. The work of \cite{2015:Morales} provides some evidence that there is a significant reduction with the result of this merge. The batch-based method can reduce the decision-making time and the management time and, in addition, can be used to the \emph{stream processing} monitoring. However, all this must be in accordance with the Lambda Architecture paradigm.

At the same time, the BIGhybrid simulator could be employed to estimate an execution time. The BIGhybrid will use the computational resources found in the environment, such as hardware performance, network performance, and tasks costs. A weight rating of computing will be defined through the simulation to define some thresholds. The restriction will aid to control overhead levels of run time. A batch method for the \emph{stream processing} enables to overcome the limitations of BIGhybrid and adapting it to low latency processing.

Estimating an approximate execution time, will make the scheduling and re-scheduling easier, because this makes it possible to know when a task will probably end at a particular computing resource before it starts. Knowing the execution time and using its heuristics will help to make a more accurate task allocation. The heuristics should include an analysis of the entire environment. However, methods such as Complex Event Processing (CEP), Event Stream Processing (ESP), Directed Acyclic Graph (DAG), as well as Machine Learning are possible features that can assist in structuring the algorithm.

\begin{figure}[htp]
\centering
\includegraphics[width=.30\textwidth]{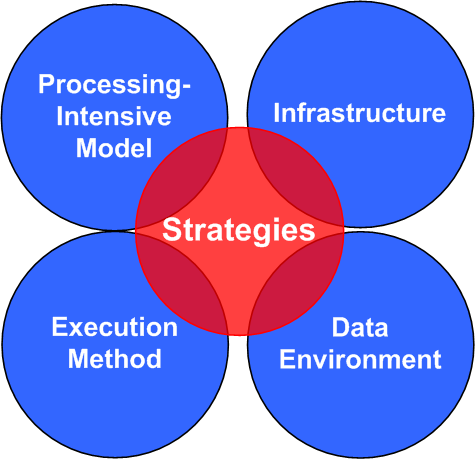}
\caption{Strategies for Reduction of Application Time}
\label{fig:strategies}
\end{figure}

Thus, if the methods mentioned above (heuristics of scheduling/rescheduling, migration, aggregation, simulation, and replication) are employed as part of the strategies for reducing the application time (that are shown in Figure  \ref{fig:strategies}), the result is also the execution time reduction. The time reducing of methods and application might achieve either a Good or the Best Result as shown in Figure \ref{fig:decision_engine_goal}.

\section{Conclusion and Future Work}
\label{sec:conclusion}
\emph{Stream processing} applied to volatile and heterogeneous environments is currently a significant subject for research. This paper describes its degree of complexity as well as current trends committed towards this subject. In response to the challenges posed by this complexity, a solution has been proposed to improve this type of data processing. The proposed solution will be applied at a complex infrastructure (i.e., geographical distributed) to study its issues and validate the model. The migration, scheduling (MapReduce simulation and heuristics), and replication features will treat the problems of its volatility, heterogeneity and dynamic. Furthermore, through the strategies to combine the desirable features, the proposed model will provide a Good or The Best Result on the scheduling settlements. Therefore, this work contributes towards a solution to volatile and heterogeneous environments for \emph{stream processing}. It will extend the SMART project and address open challenges and issues such as data stream latency, network latency, management of the computational resource, and fault control. The decisions (heuristics) and the views about the execution method (batch- or stream-based) will be a means of reducing the limitations of the system. The asynchronous and background executions will assist the required latency for performing \emph{stream processing}.

\bibliographystyle{ifacconf.bst}
\balance
\bibliography{ifacconf}             

\end{document}